# A Brief Survey on Replica Consistency in Cloud Environments

Robson A. Campêlo · Marco A. Casanova · Dorgival O. Guedes · Alberto H. F. Laender

**Abstract** Cloud computing is a general term that involves delivering hosted services over the Internet. With the accelerated growth of the volume of data used by applications, many organizations have moved their data into cloud servers to provide scalable, reliable and highly available services. A particularly challenging issue that arises in the context of cloud storage systems with geographically-distributed data replication is how to reach a consistent state for all replicas. This survey reviews major aspects related to consistency issues in cloud data storage systems, categorizing recently proposed methods into three categories: (1) fixed consistency methods, (2) configurable consistency methods and (3) consistency monitoring methods.

## 1 Introduction

Cloud computing is a general term that includes the idea of delivering hosted services over the Internet. The term "cloud" is an abstraction of this new model that arose from a common representation of a network: since the particular location of a service is not relevant, it means that services and data providers are seen as existing "in the network cloud".

Robson A. Campêlo · Dorgival O. Guedes · Alberto H. F. Laender
Department of Computer Science, Universidade Federal de Minas Gerais, 31270-901 Belo Horizonte, MG, Brazil
E-mail: {robson.campelo, dorgival, laender}@dcc.ufmg.br

Marco A.Casanova
Department of Informatics, Universidade Católica do Rio de Janeiro, 22451-900 Rio de Janeiro, RJ, Brazil
E-mail: casanova@inf.puc-rio.br

In recent years, cloud computing has emerged as a paradigm that attracts the interest of organizations and users due to its potential for cost savings, unlimited scalability and elasticity in data management. In that paradigm, users acquire computing and storage resources in a pricing model that is known as *pay-as-you-go* [8]. According to such a model, IT resources are offered in an unlimited way and the payment is made according to the actual resources used for a certain period, similarly to the traditional home utilities model.

Depending on the kind of resource offered to the users, cloud services tend to be grouped in the following three basic models: *Software as a Service* (SaaS) [30], *Platform as a Service* (PaaS) [13] and *Infrastructure as a Service* (IaaS) [16]. As an extension of this classification, when the service refers to a database, the model is known as *Database as a Service* (DBaaS) [27], which is the focus of this survey. Such a model provides transparent mechanisms to create, store, access and update databases. Moreover, the database service provider takes full responsibility for the database administration, thus guaranteeing backup, reorganization and version updates.

The use of DBaaS solutions enables service providers to replicate and customize their data over multiple servers, which can be physically separated, even placed in different datacenters [62]. By doing so, they can meet growing demands by directing users to the nearest or most recently accessed server. In that way, replication allows them to achieve features such as fast access, improved performance and higher availability. Thus, replication has become an essential feature of this storage model and is extensively exploited in cloud environments [21,41].

A particularly challenging issue that arises in the context of cloud storage systems with geographically-distributed data replication is how to reach a consistent state in all replicas. Enforcing synchronous replication to ensure strong consistency in such an environment incurs in significant performance overheads due to the increased network latency between datacenters [38] and the fact that network partitions may lead to service unavailability [19]. As a consequence, specific models have been proposed to offer weaker or relaxed consistency guarantees [60].

Several cloud storage services choose to ensure availability and performance even in the presence of network partitions rather than to offer a stronger consistency model. NoSQL-based data storage environments provide consistency properties in eventual mode [60], which means that all changes to a replicated piece of data eventually reach all its replicas. However, using this type of consistency increases the probability of reading obsolete data, since the replicas being accessed may not have received the most recent writes. This led to the development of adaptive consistency solutions, which allow adjusting the level of consistency at run-time in order to improve performance or reduce costs, while maintaining the percentage of obsolete reads at low levels [23, 32, 57].

A consistency model in distributed environments determines which guarantees can be expected for an update operation, as well as for accessing an updated object. Obtaining the correct balance between higher levels of consistency and availability is one of the open challenges in cloud computing [31]. In this survey, we focus on state-of-the-art methods for consistency in cloud environments. Considering the different solutions, we categorize such methods into three distinct categories: (1) fixed consistency methods, (2) configurable consistency methods and (3) consistency monitoring methods. Other surveys on distinct issues related to replica consistency have been recently published [6, 17, 59]. We refer the reader to them for further considerations on this topic.

The remainder of this survey is organized as follows. In Section 2, we present general concepts related to cloud database management. In Section 3, we approach the main consistency models adopted by existing distributed storage systems. In Section 4, we first propose a taxonomy to categorize the most prominent consistency methods found in the literature and then present an overview of the main approaches adopted to implement them. In Section 5, we provide a sum-up discussion emphasizing the main aspects of the surveyed methods. Finally, in Section 6, we conclude the survey by summarizing its major issues and providing some final remarks.

## 2 Cloud database management

In this section, we present general concepts related to cloud database management in order to provide a better understanding of the key issues that affect replica consistency in cloud environments. Initially, we highlight the cloud storage infrastructure requirements and describe the ACID properties [40]. Then, we introduce the CAP Theorem [19] and discuss its trade-offs.

### 2.1 Cloud data storage requirements

A trustworthy and appropriate data storage infrastructure is a key aspect to provide an adequate cloud data storage infrastructure, so that all resources can be efficiently used and shared to reduce consistency issues. Next we list some crucial requirements that must be considered by a shared infrastructure model [54].

**Automation.** The data storage must be automated to be able to quickly execute infrastructure changes required to maintain replica consistency with no human intervention.

**Availability.** The data storage must ensure that data continues to be available at a required level of performance in situations ranging from normal to adverse.

**Elasticity.** Not only must the data storage be able to scale with increasing load, but it must also be able to adjust to reductions in load by releasing cloud resources, while guaranteeing compliance with a Service Level Agreement (SLA).

**Fault tolerance.** The data storage must be able to recover in case of failure, *e.g.*, by providing a backup instance of the application that will be ready to take over without disruption.

**Low latency.** The data storage must handle latency issues by measuring and testing the network latency, before it saves the data that an application changed and before it makes such data available to other applications.

**Partition tolerance.** The data storage must be tolerant to network partitions, *i.e.*, the system must continue to operate despite them.

**Performance.** The data storage must provide an infrastructure that supports fast and robust data access, update and recovery.

**Reliability.** The data storage must ensure that the data can be recovered in case a disaster occurs.

**Scalability.** The data storage needs to quickly scale to meet workload demands, thus providing horizontal and vertical scalability. Horizontal scalability refers to the ability to increase capacity by adding more machines or setting up a new cluster or a new distributed environment. Vertical scalability, on the other hand, refers to the increase of capacity by adding more resources to a machine (*e.g.*, more memory or an additional CPU).

2.2 The ACID properties

Data Base Management Systems must conform to four transaction properties - *Atomicity*, *Consistency*, *Isolation* and *Durability* - known as the *ACID properties* [40]. However, it is non-trivial to ensure the ACID properties in a cloud data storage, exactly because data is replicated over multiple servers.

Despite this difficulty, strategies have been proposed to attempt to emulate the ACID properties for web application transactions. For instance, atomicity might be guaranteed by implementing the two-phase commit (2PC) protocol [39], whereas isolation can be obtained by a multi-version concurrency control or by a global timestamp, and durability by applying queuing strategies such as FIFO (*First-In, First-Out*) to concurrent write transactions, so that old updates do not override the latest ones [61]. However, replication represents an important obstacle to guarantee consistency [4]. Thus, maintaining a replicated database in a mutually consistent state implies that, in all replicas, each of their data items must have identical values [52]. Therefore, strategies for data update and propagation must be implemented to ensure that, if a copy is updated, all others must also be updated [56].

2.3 The CAP Theorem

The CAP Theorem was proposed by Brewer[1] as a conjecture and subsequently proved (in a restricted form) by Gilbert and Lynch [36]. Since then it has become an important concept in cloud systems [19]. It establishes that, when considering the desirable properties of *Consistency*, *Availability* and *Partition tolerance* in distributed systems, at most two of them can be simultaneously achieved.

It is evident that the CAP Theorem introduces conflicts and imposes several challenges to distributed systems and service providers. Among the conflicts, considering that network partitions are inevitable in a geographically distributed scenario, we highlight the trade-off between Consistency and Availability [37]. To illustrate this situation, in Figure 1 we observe that User 2 performs a read request for data item D1 in replica R3

[1] "Towards Robust Distributed Systems", invited presentation at the 19th Annual ACM Symposium on Principles of Distributed Computing, Portland, Oregon, July 16-19, 2000.

(Datacenter 2), after User 1 has updated data item D1 in replica R1 (Datacenter 1) in the presence of a network partition that isolates the two datacenters. Considering that the network partition means that the update made by User 1 has not been propagated to replica R3, there are two possible scenarios: the replicas may be available and User 2 will read obsolete data, thereby violating consistency, or User 2 must wait until the network partition is fixed and the update has been propagated to replica R3, thus violating availability.

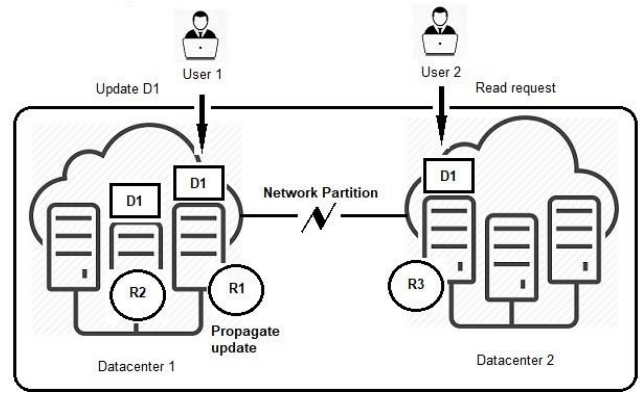

**Fig. 1** Consistency vs. Availability in Replicated Systems.

The trade-offs caused by the CAP Theorem led to the proliferation of non-ACID systems for building cloud-based applications, known as BASE [34] (systems that are *basically available*, rely on the maintenance of a *soft-state* that can be rebuilt in case of failures and are only *eventually consistent* to be able to survive network partitions). Such not-ACID systems offer distinct consistency models, which are discussed next.

3 Consistency models

A *consistency model* may be defined as a contract between a data storage system and the data processes that access it [56], thus defining strategies that support consistency within a distributed data storage system. However, trade-offs due to the CAP theorem require choosing from a range of models to address different consistency levels, which may vary from a relaxed model to a strict one [14]. In this context, there are two distinct perspectives to be considered in a distributed data storage system with respect to consistency [56]: *data-centric* and *client-centric*.

From the data-centric perspective, the distributed data storage system synchronizes the data access operations from all processes to guarantee correct results. From the client-centric perspective, the system only synchronizes the data access operations of the same process, independently from the other ones, to guarantee their consistency. This perspective is justified because it is often the case that shared updates are rare and access mostly private data.

3.1 Data-centric consistency models

In this perspective, the consistency models seek to ensure that the data access operations follow certain rules that guarantee that the storage system works correctly. These rules are based on the definition of the results that are expected after read and write operations, even considering that those operations are concurrently performed. However, the absence of a global clock makes the identification of the last write operation a difficult task, which requires some restrictions on the data values that can be returned by a read operation, thus leading to a range of consistency models. The consistency models that fall in this category are [56]: *weak consistency, PRAM consistency, causal consistency, sequential consistency* and *strict consistency*.

**Weak Consistency.** As its name indicates, weak consistency offers the lowest possible ordering guarantee, since it allows data to be written across multiple nodes and always returns the version that the system first finds. This means that there is no guarantee that the system will eventually become consistent.

**PRAM Consistency.** PRAM (Pipelined Random Access Memory) consistency, also known as FIFO consistency, is a model in which write operations from a single process are seen by the other processes in the same order that they were issued, whereas writes from different processes may be seen in a different order by different processes. In other words, there is no guarantee on the order in which the writes are seen by different processes, although writes from a single source must keep their order as if they were in a pipeline [45, 56].

**Causal Consistency.** Causal consistency is a model in which a sequential ordering is maintained only between requests that have a *causal dependency*. Two requests A and B have a causal dependency if at least one of the following two conditions is achieved: (1) both A and B are executed on a single thread and the execution of one precedes the other in time; (2) B reads a value that has been written by A. Moreover, this dependency is transitive, in the sense that, if A and B have a causal dependency, and B and C have a causal dependency, then A and C also have a causal dependency [56, 60]. Thus, in a scenario of an always-available storage system in which requests have causal dependencies, a consistency level stricter than that provided by the causal model cannot be achieved due to trade-offs of the CAP Theorem [5, 48].

**Sequential Consistency.** Sequential consistency is a stricter model that requires that: (1) all operations be serialized in the same order in all replicas; and (2) all operations from the same process be executed in the order that the storage system received them [56].

**Strict Consistency.** Strict consistency is a model that provides the strongest consistency level. It states that, if a write operation is performed on a data item, the result needs to be instantaneously visible to all processes, regardless of the replica over which the write operation was executed. To achieve that, an absolute global time order must be maintained [56].

3.2 Client-centric consistency models

In this perspective, a distributed data store is characterized by a relative absence of simultaneous updates. The emphasis is then to maintain a consistent view of data items for an individual client process that is currently operating on the data store. The consistency models that fall in this category are [56]: *eventual consistency, monotonic reads consistency, monotonic writes consistency, read-your-writes consistency* and *writes-follow-reads consistency*.

**Eventual Consistency.** This model states that all updates will propagate through the system and all replicas will gradually become consistent, after all updates have stopped for some time [56, 60]. Although this model does not provide concrete consistency guarantees, it is advocated as a solution for many practical situations [10–12, 24, 60] and has been implemented by several distributed storage systems [21, 28, 35, 43].

**Monotonic Read Consistency.** This model guarantees that if a process reads a version of a data item $d$ at time $t$, it will never see an older version of $d$ at a later time. In a scenario where data visibility is not guaranteed to be instantaneous, at least the versions of a data item will become visible in chronological order [56, 60].

**Monotonic Write Consistency.** This model guarantees that a data store must serialize two writes $w_1$ and $w_2$ in the same order that they were sent by the same client [56, 60]. For instance, if the initial write operation $w_1$ is delayed, it is not allowed for a subsequent write $w_2$ to overwrite that data item before $w_1$ completes.

**Read-Your-Writes Consistency.** This model is closely related to the monotonic read model. It guarantees that once a write operation is performed on a data item *d*, its effect will be seen by any successive read operation performed on *d* by the same process [56,60]. This means that if a client has written a version *v* of a data item *d*, it will always be able to read a version at least as new as *v*.

**Writes-Follow-Reads Consistency.** This model guarantees that if a write operation *w* is requested by a process on a data item *d*, but there has been a previous read operation *r* on *d* by the same process, then it is guaranteed that *w* will only be executed on the same or more recent value of *d* previously read. [56].

## 4 Replica consistency methods

In this section, we present an overview of state-of-the-art methods for replica consistency in cloud environments. This overview includes those methods that we considered to be among the most representative in the literature. Based on the similarities of their core ideas, we classified these methods into three distinct categories: (1) fixed consistency methods; (2) configurable consistency methods; and (3) consistency monitoring methods. In what follows, we first describe the generic characteristics of each category and then present an overview of its specific methods.

### 4.1 Fixed consistency methods

This category includes those methods that provide predefined, fixed consistency guarantees in cloud storage systems. Representative methods in this category are of two types: *Event Sequencing-based Consistency* and *Clock-based Strict Consistency*. They are described next.

#### 4.1.1 Event sequencing-based consistency

Event sequencing-based consistency methods aim at hiding replication complexity based on the fact that transaction serializability is costly and often unnecessary in web applications [11]. Thus, they provide a simple, but in many situations, effective consistency guarantee solution.

The most representative system that implements this type of method is PNUTS [25], a massively parallel and geographically distributed DBMS developed by Yahoo!. Since Yahoo!'s web applications must provide a high degree of availability for their users and must be able to read data in the presence of failures, PNUTS not only provides these features but it also supports a high degree of fault tolerance, including network partitions. PNUTS architecture is divided into regions, which contain a complete copy of each table. Thus, multiple regions containing replicated data provides additional reliability. Furthermore, PNUTS stores structured metadata in directories, which implies that users can leverage on PNUTS scalability and low latency to ensure high performance for metadata operations, such as file creation, deletion and renaming. In short, PNUTS properly manages metadata without sacrificing scalability.

PNUTS developers observed that web applications typically manipulate one record at a time, whereas different records may be located in different geographic localities. Hence, an event sequencing-based consistency method establishes that all replicas of a given record receive all updates applied to that record in the same order. This strategy is implemented by designating one of the replicas as the master for each record, so that this master receives all writes sent to that record by the other replicas. If a record has the majority of its writes sent to a particular replica, this replica becomes the master for that record.

#### 4.1.2 Clock-based strict consistency

Clock-based strict consistency methods are characterized by the use of clock-based mechanisms to control timestamps to enforce strict consistency [26, 29]. They offer the guarantee that arbitrary objects in the data store are accessed atomically and isolated from concurrent accesses. The approach behind this consistency method is based on the ability of a system to provide a timestamp log to track the order in which operations occur. According to Bravo et al. [18], this type of technique is implemented by the data storage systems themselves and might impact the consistency guarantees that they provide.

Spanner [26] and Clock-SI [29] are representative systems that implement this type of consistency method. Spanner is a scalable, globally-distributed NewSQL database service designed, built and deployed by Google. Spanner combines and extends ideas from two research communities: the database community and the systems community. From this last one, scalability and fault tolerance are the most representative features provided by Spanner. Since replication is used for global availability and geographic locality, applications can use Spanner for high availability, even in the face of wide-area natural disasters. Spanner allows different applications' data to be partitioned across different sets of servers in the same datacenter. For this reason, partition tolerance is an important requirement in Spanner. In addition, Spanner provides constraints for controlling read and write latency. Spanner also assigns globally-meaningful commit timestamps that reflect the serialization order of the transactions, which may be distributed. Moreover, Spanner enforces that if a transaction $T_2$ begins after the commit of a transaction $T_1$, then the commit timestamp of $T_2$ must be greater than the commit timestamp of $T_1$.

On the other hand, Clock-SI provides a fully distributed protocol for partitioned data stores that supports availability and scalability, bringing performance benefits. It also avoids a "single point of failure" and, therefore, a potential performance bottleneck, thus improving transaction latency and throughput. Clock-SI implements the so called *snapshot isolation* replication, which is a consistency criterion for partitioned data storage. In this strategy, read-only operations read from a consistent snapshot and other operations perform a commit if no objects written by these transactions were concurrently written. The local physical clock is used by each transaction to identify its read timestamp.

Another example of a system that implements snapshot isolation replication is Vela [55], which is a system for running off-the-shelf relational databases on the cloud. Vela provides a primary master and two secondary replicas, which are synchronously replicated, thereby offering fault tolerance. In addition, Vela relies on hardware virtualization to improve performance by reducing complexity with a minimal cost in resources. Instead of using data replication for durability, Vela uses this technique for improving latency, decoupling the update and the read-only workloads. Moreover, Vela provides elasticity, by monitoring the CPU idle time percentage, and scalability, by adding new replicas according to its workload.

### 4.2 Configurable consistency methods

The methods included in this category implement mechanisms that provide configurable levels of consistency, such as self-adaptive or flexible consistency guarantees, which allow the selection or specification of the desired consistency level at any given point. These methods are of two types, *Automated and Self-adaptive Consistency* and *Flexible Consistency*, and are described next.

#### 4.2.1 Automated and self-adaptive consistency

Automated and self-adaptive consistency methods aim at dynamically enforcing multiple consistency degrees over distinct data objects. Three main approaches have been adopted to implement them.

**Stale reads estimation**. This approach is based on an estimation of the rate of read operations that return instances of data objects that have already been updated to newer values - the *stale reads* [47]. Once the estimation model is computed, it is possible to identify the key parameter that affects the stale reads and then to scale up/down the number of replicas. This approach is mainly implemented by Harmony [23], which is a cloud storage system that automatically identifies the key parameters affecting the stale reads, such as system states and application requirements. Harmony adds minimal latency while reducing the stale data reads by almost 80%. Its goal is to gradual and dynamically tune the consistency level at run time according to the applications' consistency requirements, in order to provide adequate tradeoffs between consistency and both performance and availability. An intelligent estimation model of stale reads is the key aspect of Harmony. Its mechanism of elastically scaling up/down the number of replicas maintains a minimal tolerable fraction of stale reads, which results on meeting the required level of consistency while achieving good performance.

**Divergence bounds enforcement**. This approach allows the evaluation and enforcement of divergence bounds over data objects (table/row/column), thus providing consistency levels that can be automatically adjusted based on statistical information [53]. The evaluation takes place on a divergence vector every time an update request is received, although it is necessary to identify the affected data objects. If any limit is exceeded, all updates since the last replication are placed in a FIFO-like queue to be propagated and executed on the other replicas. The most representative system that implements this approach is VFC$^3$ (*Versatile Framework for Consistency in Cloud Computing*) [32], which adopts a consistency model for replicated data across datacenters. The VFC$^3$ model considers the different data semantics and automatically adjusts the consistency levels based on statistical information. Its main goal is to offer control over consistency to provide high-availability without compromising performance. Furthermore, VFC$^3$ targets cloud tabular data stores, offering rationalization of resources and improvement of Quality-of-Service (QoS), thereby reducing latency.

**Dynamic allocation**. This approach dynamically selects to which server (or even a set of servers) each read of a data item must be directed, so that the best service is delivered given the current configuration and system conditions. Hence, this approach is adaptable to distinct configurations of replicas and users, as well as to changing conditions, such as variations on the network performance or server load. Another important aspect of this approach is the fact that it allows application developers to provide a Service Level Agreement (SLA) that specifies the applications' consistency/latency desires in a declarative manner. Pileus [57] is a key-value storage system that implements this approach by providing a diversity of consistency guarantee options for globally distributed and replicated data environments. In fact, Pileus allows several systems or clients of a system to achieve different consistency degrees, even when they share the same data. Moreover, Pileus supports availability and performance by limiting the set of suitable servers, whereby strong reads must be directed to the primary site and eventual reads can be answered by any replica. A large table can be split into one or more smaller tables in order to achieve scalability. In Pileus, users perform operations to access data that is partitioned and replicated among distinct servers, thus it must support partition tolerance and reliability.

*4.2.2 Flexible consistency*

Flexible consistency methods cover distinct approaches that adapt to predefined consistency models in a flexible way. There are four main approaches that implement such methods.

**Invariants-based**. This approach strengthens eventual consistency, thus allowing the applications to specify consistency rules, or *invariants*, that must be maintained by the system. Once those invariants are defined, it is possible to identify the operations that are potentially unsafe under concurrent execution, thus allowing one to select either a violation-avoidance or an invariant-repair technique. Indigo [12] is a middleware system that implements this approach. It supports an alternative consistency method built on top of a geo-replicated and partitioned key-value data store. In addition, Indigo guarantees strong application invariants, while providing low latency to an eventually-consistent system. Indigo builds on the fault tolerance of the underlying storage system, thereby the failure of a machine inside a datacenter does not lead to any data loss.

**Linearizable/eventual**. This approach supports updates with a choice of linearizable and eventual consistency. GEO [15] is an open-source geo-distributed actor[2] system that implements this approach. It supports both replicated and single-instance coherence protocols. Replication can provide fast, always-available reads and updates. GEO also improves performance by caching actor states in one or more datacenters. Furthermore, the geo-distributed actor system can reduce access latencies by exploiting locality, since the caching policy for each actor can be declared as single-instance or multi-instance. Caching multiple instances can reduce the access latency for actors without locality. In GEO, an actor may be declared as *volatile* or *persistent*. In the first case, the latest version resides in memory and may be lost when servers fail, whereas in the second case the latest version resides in the storage layer. In case a user requests linearizability, GEO guarantees true linearizability in real time, between call and return.

---

[2] Service applications can use actors to provide a programming model to simplify synchronization, fault-tolerance and scalability. It represents a useful abstraction for the middle tier of scalable service applications that run on a virtualized cloud infrastructure in a datacenter [15].

**Requirements-based**. This approach supports the applications' consistency requirements. The discussion that follows adopts the notion of service (requested by an application) as a generalization of the read and write operations used thus far. The trade-off between consistency and scalability requirements is handled by introducing the notion of consistency regions and service-delivery-oriented consistency policies. A consistency region is defined as a logical unit that represents the application state-level requirements for consistency and scalability. This concept is used to define consistency boundaries that separate each group of services that need to be ordered. Hence, services that need to be kept consistent must be associated to a certain region. The definition of what region a service belongs to and which services that can be concurrently delivered is set by the system administrators. Scalable Service Oriented Replication (SSOR) [22] is a middleware that implements this approach. SSOR presents a Region-based Election Protocol (REP) that provides a mechanism to balance the workload amongst sequencers, thereby efficiently improving elasticity in the cloud. Replication is used by SSOR to provide end-to-end fault tolerance between end clients and cloud services. In order to ensure reliability, SSOR implements solutions that tolerate sequencers and nonsequencers crashes, and introduces the concept of region distribution synchrony for handling simultaneous node crashing. A better performance is also achieved by reducing the load on the sequencer, extending the Multi-fixed Sequencers Protocol (MSP).

SSOR covers three distinct types of consistency region: (1) Conflict Region (CR), which is a region composed by services that have conflicting requirements for consistency regardless of the session; (2) Sessional Conflict Region (SCR), which is a region that includes services of a particular session with conflicting consistency requirements; and (3) Non-Conflict Region (NCR), which is a region that does not impose any consistency constraints or requirements.

**Adaptable**. This approach handles unpredictable workloads by allowing the system to be tuned for capacity in an elastic and flexible way. Due to this characteristic, it allows applications to perform eventually or strongly consistent reads as needed. Amazon DynamoDB[3] [2] is a highly reliable and cost-effective NoSQL database service that implements this approach. DynamoDB provides high availability and high throughput at very low latency. It has been designed to be scalable and to achieve high performance even at high scale. It was built based on the experience with its predecessor Dynamo [28]. DynamoDB adopts eventual consistency as its default model, which does not guarantee that an eventually consistent read will always reflect the result of a recently completed write. On the other hand, when adopting a stronger consistency model, it returns a result that reflects all writes that have received a successful response prior to that read.

### 4.3 Consistency monitoring methods

Alternatively, instead of directly handling data consistency issues, some methods focus on providing mechanisms that allow data owners to detect the occurrence of consistency violations in the cloud storage. This means that clients might audit their own data and make decisions based on how the Cloud Service Provider (CSP) stores and manages their replicas according to the consistency level that has been agreed upon in the service level contract. The methods in this category are of two types, *Consistency Verification* and *Consistency Auditing*, and are described next.

#### 4.3.1 Consistency verification

Consistency verification methods are based on two approaches, namely *protocol-based* and *contract-based*.

**Protocol-based**. This approach is based on a protocol that enables a group of mutually trusting clients to detect consistency violations on a cloud storage. It is adopted by VICOS (Verification of Integrity and Consistency for Cloud Object Storage) [17]. VICOS supports the concept of *fork-linearizability*, which captures the strongest achievable notion of consistency in multi-client models. The method may guarantee this notion by registering the causal evolution of the user's views into their interaction with the server. When the server creates only a single discrepancy between the views of two clients, it is ensured that these clients will never observe each other's operations afterwards. That is, if these users later communicate and the server lies to them, the violation will be immediately discovered. Thus, users can verify a large number of past transactions by performing a single check.

**Contract-based**. This approach provides a verification scheme that allows data owners to ensure whether the CSP complies with the SLA for storing data in multiple replicas. It is implemented by DMR-PDP (*Dynamic Multi-Replica Provable Data Possession*) [51]. The context addressed by this scheme is that whenever data owners ask the CSP to replicate data at different servers, they are charged for this. Hence, data owners need to be strongly persuaded that the CSP stores all data copies that are agreed upon in the service level contract, as well as that all remotely stored copies correctly execute the updates requested by the users. This approach deals with such problems by preventing the CSP from cheating the data storage, for instance, by maintaining fewer copies than paid for. Such scheme is based on a technique called *Provable Data Possession* [9], which is used to audit and validate the integrity and consistency of data stored on remote servers.

#### 4.3.2 Consistency auditing

This type of method is based on an architecture that consists of a large data cloud maintained by a CSP and multiple small audit clouds composed of a group of users that cooperate on a specific job (*e.g.*, revising a document or writing a program). The required level of consistency that should be provided by the data cloud is stipulated by an SLA involving the audit cloud and the data cloud. Once the SLA is defined, the audit cloud can verify whether the data cloud violates it, thus quantifying, in monetary terms or otherwise, the severity of the violation.

Consistency as a Service (CaaS) [46,49] implements this method. It relies on a two-level auditing structure, namely: *local auditing* and *global auditing*. Local auditing allows each user to independently perform local tracing operations, focusing on monotonic read and read-your-write consistencies. Global auditing, on the other hand, requires that an auditor be periodically elected from the audit cloud to perform global tracing operations, focusing on causal consistency. This method is supported by constructing a directed graph of operations, called the *precedence graph*. If the graph is acyclic, the required level of consistency is preserved [46].

## 5 Discussion

As proposed in Section 4, replica consistency methods can be grouped in three categories: fixed consistency methods, configurable consistency methods and consistency monitoring methods.

---

[3] https://www.allthingsdistributed.com/2012/01/amazon-dynamodb.html

**Table 1** Summary of the Surveyed Replica Consistency Methods

| Category | Method | Brief Description |
| --- | --- | --- |
| Fixed Consistency | Event Sequencing-based Consistency [25] | Establishes that all replicas of a given record apply all updates to a record in the same order and is, therefore, related to **sequential consistency**. |
| | Clock-based Strict Consistency [26, 29, 55] | Uses clock-based mechanisms to control timestamps to enforce **strict consistency**. |
| Configurable Consistency | Automated and Self-Adaptive Consistency [23, 32, 57] | Provides a gradually and dynamically **tunable consistency** at runtime according to the applications' consistency requirements. Enforces **increasing degrees of consistency** for different types of data, based on their semantics. |
| | Flexible Consistency Guarantees [2, 12, 15, 22] | Allows applications to specify **consistency rules**, or **invariants**, that must be maintained by the system. Supports updates with a choice between **linearizable consistency** and **eventual consistency**. Supports the **applications' consistency requirements** and flexibly adapt to **predefined consistency models**. Allows applications to perform **eventually or strongly consistent reads** as needed. |
| Consistency Monitoring | Consistency Verification [17, 51] | Enables a group of mutually trusting clients to detect **data-integrity** and **consistency violations**. Allows the data owner to ensure that the Cloud Service Provider stores all data copies that are agreed upon in the **service level contract.** |
| | Consistency Auditing [46, 49] | Implements a Local and Global Auditing structure to allow a group of clients to detect **consistency violations**. |

Fixed consistency methods are mostly based on versioning of events. They capture the idea of event ordering by means of control strategies such as a sequence number that represents a data object version or clock-based mechanisms which are well-understood concepts in distributed systems [33,44,50]. The idea of an event happening before another represents a causal relationship and the total ordering of events among the replicas has been shown quite useful for solving synchronization issues related to data consistency. Thus, consistency methods in this category extend this concept on specific scenarios.

Configurable consistency methods, in turn, generally aim at providing mechanisms that automatically adjust the degree of consistency. This is an important feature for applications that have temporal characteristics, as well as for real-time workload cloud storage systems. Specifically, configurable consistency methods are suitable to address applications' consistency requirements that need to adapt to predefined consistency models.

On the other hand, consistency monitoring methods do not provide specific guarantees, but focus on detecting the occurrence of consistency violations in the cloud data storage. Despite that, these methods offer significant contributions that are suitable for scenarios where multiple clients cooperate on remotely stored data in a potentially misbehaving service and need to rely on the CSP to guarantee their correctness. Furthermore, those clients need to verify if the requested data updates were correctly executed on all remotely stored copies, while maintaining the required consistency level.

Table 1 summarizes the surveyed methods. In particular, the description column briefly relates the consistency models of Section 3 with the methods addressed in Section 4 (keyword terms shown in boldface). The reader is thereby implicitly invited to compare the methods based on the characteristics of the consistency models they support. The relationships are not entirely crispy, though, since some methods are flexible with respect to the consistency model they follow, whereas others are application-dependent, based on a contract or service level agreement between the application and the system.

Table 2 summarizes the storage requirements supported by the systems that we have addressed in order to stress what are the main consistency trade-offs they consider, in the broad perspective of the CAP Theorem. Note that in Table 2 we only address those systems that implement a specific consistency method, since consistency monitoring methods only focus on detecting consistency violations.

As previously mentioned in Section 2.3, the existing trade-off determined by the CAP Theorem implies that applications must sacrifice consistency to be able to satisfy other application requirements. Thus, Table 2 shows Availability and Partition Tolerance as storage requirements, which are related to the CAP Theorem. In regard to the remaining requirements in Table 2, they are not directly related to the CAP Theorem, but have some impact on the consistency methods (See Sect. 2.1).

**Table 2** Storage Requirements Supported by the Fixed and Configurable Replica Consistency Methods

| Category | Representative Systems | Automation | Availability | Elasticity | Fault Tolerance | Low Latency | Partition Tolerance | Performance | Reliability | Scalability |
|---|---|---|---|---|---|---|---|---|---|---|
| Fixed Replica Consistency Methods | PNUTS [25] |  | √ |  | √ | √ | √ | √ | √ | √ |
|  | Spanner [26] |  | √ |  | √ | √ | √ |  |  | √ |
|  | Clock-SI [29] |  | √ |  |  | √ | √ | √ |  | √ |
|  | Vela [55] |  |  | √ | √ | √ |  | √ |  | √ |
| Configurable Replica Consistency Methods | Indigo [12] |  |  |  | √ | √ | √ |  |  |  |
|  | GEO [15] |  | √ |  | √ | √ |  | √ |  | √ |
|  | SSOR [22] |  |  | √ | √ |  |  | √ | √ |  |
|  | Harmony [23] | √ | √ | √ |  | √ |  | √ |  |  |
|  | VFC [32] | √ | √ |  |  | √ |  | √ |  |  |
|  | DynamoDB [2] |  | √ |  |  | √ | √ | √ | √ | √ |
|  | Pileus [57] | √ | √ |  |  |  | √ | √ | √ | √ |

Table 2 shows that PNUTS [25], Spanner [26], Clock-SI [29], DynamoDB [2] and Pileus [57] guarantee at the same time Availability and Partition Tolerance, thus providing a weaker type of consistency. Therefore, these systems address the CAP Theorem trade-offs by sacrificing consistency.

PNUTS, Spanner and Clock-SI implement fixed consistency methods, which means that the consistency criteria these systems adopt are not flexible. On the other hand, DynamoDB and Pileus implement configurable consistency methods, thereby tuning the consistency level is not a problem under certain scenarios. For the remaining systems, consistency is not impacted by the CAP Theorem trade-offs. Vela [55], GEO [15], Harmony [23] and VFC [32] do not guarantee partition tolerance, meaning that availability and consistency are likely to be achieved. Similarly, Indigo [12] does not guarantee availability, thus achieving consistency and partition tolerance. Finally, SSOR [22] does not guarantee availability and partition tolerance, which means that its method concerns only in providing consistency.

Although the CAP theorem addresses an important issue in distributed systems, there are other consistency-related tradeoffs that have a direct impact on modern distributed database management systems (DDBMS). These tradeoffs are particularly related to performance, scalability and latency, as described next.

**Consistency vs. performance.** As stressed by Brewer, twelve years after proposing his CAP theorem [19], the tradeoff between consistency and performance is even more relevant. He argues that partitions are rare, so that a DDBMS should consider the tradeoff consistency versus availability only when partition tolerance is required. However, the tradeoff between consistency and performance is permanent.

**Consistency vs. scalability.** A scalable system, in turn, is non-trivially achieved when consistency is required, since this can be very expensive. Thus, in order to improve scalability a relaxed consistency state is often provided by some systems. However, the price is that the state of each replica may not be always the same [22].

**Consistency vs. latency.** According to Abadi [5], there is a connection between latency (understood as the time to initiate an operation) and availability. A system becomes unavailable in the presence of high latency. On the other hand, if latency decreases, the system becomes more available. However, despite this apparently obvious implication, a system may be available, but might exhibit high latency rates. Hence, he argues that the consistency versus latency and consistency versus availability tradeoffs are connected and exist beyond the CAP theorem.

In this context, Table 2 also shows the surveyed systems in the perspective of the above tradeoffs. As we can see, performance, scalability and low latency are important requirements supported by the fixed and configurable consistency methods. However, there are relevant differences about how the respective methods handle these requirements.

Configurable consistency methods focus in general on tuning the consistency level or offering control in order to provide consistency without affecting other requirements. For instance, the stale reads estimation approach (Sec. 4.2.1) reduces the probability of stale reads caused by the cloud system dynamicity and the application's demands. Once the number of replicas involved in read operations is elastically scaled up/down to maintain a low (or zero) tolerable fraction of stale reads, this automated and self-adaptive approach provides an appropriate balance between consistency, performance and availability.

In turn, for those systems that implement fixed consistency methods, performance, scalability and latency can be achieved by using techniques such as synchronous replication. For instance, Vela uses snapshot isolation replication for achieving performance and scalability without sacrificing consistency [55]. However, the system relies on a weaker consistency level to avoid tradeoffs between consistency and performance/scalability. Much the same way, although PNUTS provides an asynchrony model of current requests for achieving low latency, thus relying on a relaxed consistency model that avoids a tradeoff with latency, it prevents increasing the consistency level [25]. Therefore, configurable consistency methods are more convenient for applications that require flexible consistency levels on demand, but that at the same time try to avoid impacting storage requirements.

## 6 Conclusions

In this survey we have reviewed several methods proposed in the literature to ensure replica consistency in distributed cloud data storage systems. Ensuring consistency in replicated databases is an important research topic that offers many challenges in the sense that such systems must provide a consistent state of all replicas, despite the occurrence of concurrent transactions. In other words, such systems must provide a suite of strategies for data update and propagation to guarantee that, if one copy is updated, all others must be updated as well. The taxonomy presented in Section 4 provides researchers and developers with a framework to better understand the current main ideas and challenges in this area.

This survey, by necessity, does not exhaust all topics related to replica consistency in distributed cloud data storage systems. As examples, we conclude with remarks about two topics not addressed in the survey, consistency recovery and trust.

**Consistency Recovery.** This topic refers to the question of how to recover from a consistency violation and is therefore directly related to the focus of the survey. An approach to address this issue in traditional systems is to introduce *compensatory actions* that restore consistency. This approach leads to the concept of *long transactions* or *sagas*, that is, sequences of transactions that, together, preserve consistency. A long transaction may accommodate consistency checks and compensatory actions among its sub-transactions. A familiar example is how airlines handle overbooking. The question then is if sagas can be generalized, or adapted, to the context of a database service provider. The IBM Cloud Functions with Action Sequences [1] and AWS Step Functions [3], for example, offer methods to connect multiple functions into a single service, but they do not focus on replica consistency issue [7].

**Trust.** This topic can be divided into three related questions: (1) May the clients of a database service provider trust the service? (2) May the service trust a client? (3) May a client trust the other clients of the database service provider? The first question is indeed an issue since a database service provider is a complex system, with many layers, that may be vulnerable to integrity and confidentiality threats. The second question is not new, but it is again exacerbated in a database service provider, given the complexity of the system. This issue is addressed, for example, by Cachin and Ohrimenko [20], as well as by Krahn et al. [42]. The third question is closely related to the previous two, but somewhat more subtle. Since a database service provider typically maximizes sharing its resources among multiple clients, the service must guarantee that malicious clients will not temper with data of the other clients. Ideally, the database service provider should maximize the number of concurrent clients, irrespectively of their level of trustworthiness. Furthermore, it should differentiate between trusted and malicious users, and assign data resources in such a way that clients in one class do not share data resources with clients in the other. This question was addressed, for example, by Thakur and Bresli [58] for Cloud service providers in general, but it remains an issue that database service providers should specifically address, perhaps by re-interpreting some concepts motivated by the replica consistency problem, such as region and snapshot isolation.

## Funding

This research was funded by the authors' individual grants from CAPES, CNPq, FAPEMIG and FAPERJ.

## References

1. Cloud Functions - Overview - IBM Cloud. [Online; accessed last time on 30-September-2019].
2. Amazon DynamoDB Developer Guide, 2012. [Online; accessed last time on 30-September-2019].
3. AWS Step Functions Developer Guide, 2019. [Online; accessed last time on 30-September-2019].
4. D. J. Abadi. Data Management in the Cloud: Limitations and Opportunities. *IEEE Data Engineering Bulletin*, 32:3–12, 2009.
5. D. J. Abadi. Consistency Tradeoffs in Modern Distributed Database System Design: CAP is Only Part of the Story. *IEEE Computer*, 45(2):37–42, 2012.
6. D. Agrawal, A. El Abbadi, and K. Salem. A Taxonomy of Partitioned Replicated Cloud-based Database Systems. *IEEE Data Engineering Bulletin*, 38(1):4–9, 2015.
7. Istemi Ekin Akkus, Ruichuan Chen, Ivica Rimac, Manuel Stein, Klaus Satzke, Andre Beck, Paarijaat Aditya, and Volker Hilt. SAND: Towards High-performance Serverless Computing. In *Proceedings of the 2018 USENIX Annual Technical Conference*, pages 923–935, Berkeley, CA, USA, 2018. USENIX Association.
8. M. Al-Roomi, S. Al-Ebrahim, S. Buqrais, and I. Ahmad. Cloud Computing Pricing Models: A Survey. *International Journal of Grid and Distributed Computing*, 6(5):93–106, 2013.
9. G. Ateniese, R. Burns, R. Curtmola, J. Herring, L. Kissner, Z. Peterson, and D. Song. Provable Data Possession at Untrusted Stores. In *Proceedings of the 14th ACM Conference on Computer and Communications Security*, pages 598–609, Alexandria, VA, USA, 2007.
10. P. Bailis, S. Venkataraman, M. J. Franklin, J. M. Hellerstein, and I. Stoica. Probabilistically Bounded Staleness for Practical Partial Quorums. *Proceedings of the VLDB Endowment*, 5(8):776–787, 2012.
11. P. Bailis, S. Venkataraman, M. J. Franklin, J. M. Hellerstein, and I. Stoica. Quantifying Eventual Consistency with PBS. *Communications of the ACM*, 57(8):93–102, 2014.
12. V. Balegas, S. Duarte, C. Ferreira, R. Rodrigues, N. Preguiça, M. Najafzadeh, and M. Shapiro. Putting Consistency Back into Eventual Consistency. In *Proceedings of the Tenth European Conference on Computer Systems*, pages 6:1–6:16, Bordeaux, France, 2015.
13. D. Beimborn, T. Miletzki, and D. I. S. Wenzel. Platform as a Service (PaaS). *Wirtschaftsinformatik*, 53(6):371–375, 2011.
14. D. Bermbach and J. Kuhlenkamp. Consistency in Distributed Storage Systems: An Overview of Models, Metrics and Measurement Approaches. In *Proceedings. of the First International Conference on Networked Systems*, pages 175–189. Marrakech, Morocco, 2013.
15. P. A. Bernstein, S. Burckhardt, S. Bykov, N. Crooks, J. M. Faleiro, G. Kliot, A. Kumbhare, M. R. Rahman, V. Shah, A. Szekeres, and J. Thelin. Geo-Distribution of Actor-Based Services. *Proceedings of the ACM on Programming Languages*, 1:107:1–107:26, 2017.
16. S. Bhardwaj, L. Jain, and S. Jain. Cloud computing: A study of infrastructure as a service (IAAS). *International Journal of Engineering and Information Technology*, 2(1):60–63, 2010.
17. M. Brandenburger, C. Cachin, and N. Knezevic. Don't Trust the Cloud, Verify: Integrity and Consistency for Cloud Object Stores. *ACM Transactions on Privacy and Security*, 20:8:1–8:30, 2017.
18. M. Bravo, N. Diegues, J. Zeng, P. Romano, and L. Rodrigues. On the use of Clocks to Enforce Consistency in the Cloud. *IEEE Data Engineering Bulletin*, 38(1):18–31, 2015.
19. Eric Brewer. Pushing the CAP: Strategies for Consistency and Availability. *IEEE Computer*, 45(2):23–29, 2012.
20. Christian Cachin and Olga Ohrimenko. Verifying the consistency of remote untrusted services with conflict-free operations. *Information and Computation*, 260:72–88, 2018.
21. F. Chang, J. Dean, S. Ghemawat, W. C. Hsieh, D. A. Wallach, M. Burrows, T. Chandra, A. Fikes, and R. E. Gruber. Bigtable: A Distributed Storage System for Structured Data. *ACM Transactions on Computer Systems*, 26(2):4:1–4:14, 2008.
22. T. Chen, R. Bahsoon, and A. H. Tawil. Scalable service-oriented replication with flexible consistency guarantee in the cloud. *Information Sciences*, 264:349–370, 2014.
23. H. Chihoub, S. Ibrahim, G. Antoniu, and M. S. Perez. Harmony: Towards Automated Self-adaptive Consistency in Cloud Storage. In *Proceedings of the 2012 IEEE International Conference on Cluster Computing*, pages 293–301, Beijing, China, 2012.
24. H. Chihoub, S. Ibrahim, G. Antoniu, and M. S. Pérez. Consistency Management in Cloud Storage Systems. In M. Gaber S. Sakr, editor, *Large Scale and Big Data: Processing and Management*. CRC Press, 2014.
25. B. F. Cooper, R. Ramakrishnan, U. Srivastava, A. Silberstein, P. Bohannon, H. Jacobsen, N. Puz, D. Weaver, and R. Yerneni. PNUTS: Yahoo!'s Hosted Data Serving Platform. *Proceedings of the VLDB Endowment*, 1(2):1277–1288, 2008.
26. J. C. Corbett, J. Dean, and M. Epstein et al. Spanner: Google's Globally Distributed Database. *ACM Transactions on Computer Systems*, 31(3):8:1–8:22, 2013.
27. C. Curino, E. Jones, R. Popa, N. Malviya, E. Wu, S. Madden, H. Balakrishnan, and N. Zeldovich. Relational Cloud: A Database-as-a-Service for the Cloud. In *Proceedings of the 5th Biennial Conference on Innovative Data Systems Research*, pages 235–240, Asilomar, CA, USA, 2011.
28. G. DeCandia, D. Hastorun, M. Jampani, G. Kakulapati, A. Lakshman, A. Pilchin, S. Sivasubramanian, P. Vosshall, and W. Vogels. Dynamo: Amazon's Highly Available Key-value Store. *ACM SIGOPS Operating Systems Review*, 41(6):205–220, 2007.
29. J. Du, S. Elnikety, and W. Zwaenepoel. Clock-SI: Snapshot Isolation for Partitioned Data Stores Using Loosely Synchronized Clocks. In *Proceedings of the IEEE 32nd International Symposium on Reliable Distributed Systems*, pages 173–184, Braga, Portugal, 2013.
30. A. Dubey and D. Wagle. Delivering software as a service. *The McKinsey Quarterly*, 6:1–7, 2007.
31. M. M. Elbushra and J. Lindström. Eventual Consistent Databases: State of the Art. *Open Journal of Databases*, 1(1):26–41, 2014.
32. S. Esteves, J. Silva, and L. Veiga. Quality-of-service for Consistency of Data Geo-replication in Cloud Computing. In *Proceedings of the 18th European Conference on Parallel Processing*, pages 285–297, Rhodes Island, Greece, 2012.
33. C. Fidge. Logical Time in Distributed Computing Systems. *IEEE Computer*, 24(8):28–33, 1991.
34. A. Fox, S. D. Gribble, Y. Chawathe, E. A. Brewer, and P. Gauthier. Cluster-based Scalable Network Services. In *Proceedings of the 16th ACM Symposium on Operating*

*Systems Principles*, pages 78–91, New York, NY, USA, 1997.

35. S. Ghemawat, H. Gobioff, and S. Leung. The Google File System. *ACM SIGOPS Operating Systems Review*, 37(5):29–43, 2003.
36. S. Gilbert and N. Lynch. Brewer's Conjecture and the Feasibility of Consistent, Available, Partition-tolerant Web Services. *ACM SIGACT News*, 33(2):51–59, 2002.
37. S. Gilbert and N. Lynch. Perspectives on the CAP theorem. *IEEE Computer*, 45(2):30–36, 2012.
38. S. Goel and R. Buyya. Data replication strategies in wide-area distributed systems. In R.G. Qiu, editor, *Enterprise Service Computing: From Concept to Deployment*, pages 211–241. IGI Global, 2007.
39. Jim Gray. Notes on Data Base Operating Systems. In *Operating Systems, An Advanced Course*, pages 393–481, London, UK, UK, 1978. Springer-Verlag.
40. T. Haerder and A. Reuter. Principles of Transaction-oriented Database Recovery. *ACM Computing Surveys*, 15(4):287–317, 1983.
41. S. Ibrahim, H. Jin, L. Lu, B. He, G. Antoniu, and S. Wu. Maestro: Replica-Aware Map Scheduling for MapReduce. In *Proceedings of the 12th IEEE/ACM International Symposium on Cluster, Cloud and Grid Computing*, pages 435–442, Ottawa, Canada, 2012.
42. Robert Krahn, Bohdan Trach, Anjo Vahldiek-Oberwagner, Thomas Knauth, Pramod Bhatotia, and Christof Fetzer. Pesos: Policy Enhanced Secure Object Store. In *Proceedings of the Thirteenth EuroSys Conference*, New York, NY, USA, 2018. ACM.
43. A. Lakshman and P. Malik. Cassandra: A Decentralized Structured Storage System. *ACM SIGOPS Operating Systems Review*, 44(2):35–40, 2010.
44. L. Lamport. Time, Clocks, and the Ordering of Events in a Distributed System. *Communications of the ACM*, 21(7):558–565, 1978.
45. R. J. Lipton and J. S. Sandberg. *PRAM: A Scalable Shared Memory*. Tecnical Report TR-180-88, Department of Computer Science, Princeton University, 1988.
46. Q. Liu, G. Wang, and J. Wu. Consistency as a Service: Auditing Cloud Consistency. *IEEE Transactions on Network and Service Management*, 11(1):25–35, 2014.
47. H. Lu, K. Veeraraghavan, P. Ajoux, J. Hunt, Y. J. Song, W. Tobagus, S. Kumar, and W. Lloyd. Existential Consistency: Measuring and Understanding Consistency at Facebook. In *Proceedings of the 25th Symposium on Operating Systems Principles*, pages 295–310, 2015.
48. P. Mahajan, L. Alvisi, and M. Dahlin. *Consistency, Availability, and Convergence*. Tecnical Report UTCS TR-11-22, Department of Computer Science, The University of Texas at Austin, 2011.
49. N. R. Math and V. Biradar. Consistency as a Service: Maintaining Cloud Consistency Using Auditing. *International Journal of Innovative Science Engineering and Technology*, 2(6):577–588, August 2015.
50. F. Mattern. Virtual Time and Global States of Distributed Systems. *Parallel and Distributed Algorithms*, 1(23):215–226, 1989.
51. R. Mukundan, S. Madria, and M. Linderman. Replicated Data Integrity Verification in Cloud. *IEEE Data Engineering Bulletin*, 35(4):55–64, 2012.
52. M. T. Özsu and P. Valduriez. *Principles of Distributed Database Systems, Third Edition*. Springer, 2011.
53. S. P. Phansalkar and A. R. Dani. Tunable consistency guarantees of selective data consistency model. *Journal of Cloud Computing: Advances, Systems and Applications*, 4(13), 2015.
54. B. P. Rimal, E. Choi, and I. Lumb. A Taxonomy and Survey of Cloud Computing Systems. In *Proceedings of the 2009 Fifth International Joint Conference on INC, IMS and IDC*, pages 44–51, Seoul, Korea, 2009.
55. T. Salomie and G. Alonso. Scaling Off-the-Shelf Databases with Vela: An Approach based on Virtualization and Replication. *IEEE Data Engineering Bulletin*, 38(1):58–72, 2015.
56. A. S. Tanenbaum and M. V. Steen. *Distributed Systems: Principles and Paradigms*. Prentice-Hall, Upper Saddle River, New Jersey, 2007.
57. D. B. Terry, V. Prabhakaran, and R. Kotla et al. Consistency-based Service Level Agreements for Cloud Storage. In *Proceedings of the Twenty-Fourth ACM Symposium on Operating Systems Principles*, pages 309–324, New York, NY, USA, 2013.
58. Subhasis Thakur and John G. Breslin. A robust reputation management mechanism in the federated cloud. *IEEE Transactions on Cloud Computing*, 7(1):625 – 637, 2019.
59. P. Viotti and M. Vukolić. Consistency in Non-transactional Distributed Storage Systems. *ACM Computing Surveys*, 49(1):19, 2016.
60. W. Vogels. Eventually Consistent. *Communications of the ACM*, 52(1):40–44, 2009.
61. Z. Wei, G. Pierre, and C. Chi-Hung. Scalable Transactions for Web Applications in the Cloud. In *Proceedings of the 15th International Euro-Par Conference on Parallel Processing*, pages 442–453, Delft, The Netherlands, 2009.
62. P. Xiong, Y. Chi, and S. Zhu et al. Intelligent Management of Virtualized Resources for Database Systems in Cloud Environment. In *Proceedings of the 27th International Conference on Data Engineering*, pages 87–98, Washington, DC, USA, 2011.